\documentclass[12pt,a4paper]{article}
\usepackage[latin9]{inputenc}
\usepackage{amsmath}
\usepackage{amssymb}
\usepackage{graphicx}

\makeatletter

\pdfpageheight\paperheight
\pdfpagewidth\paperwidth

\usepackage{jheppub}
\author[a,b,c]{R.N. Lee}
\affiliation[a]{Budker Institute of Nuclear Physics, 630090, Novosibirsk, Russia}
\affiliation[b]{Novosibirsk State University, 630090, Novosibirsk, Russia}
\affiliation[c]{Institut f\"ur Theoretische Teilchenphysik Karlsruhe Institute of Technology (KIT)
76128 Karlsruhe, Germany}
\author[d,2]{V.A. Smirnov}%
\affiliation[d]{Skobeltsyn Institute of Nuclear Physics of Moscow State University, 119992 Moscow, Russia}
\emailAdd{R.N.Lee@inp.nsk.su}
\emailAdd{smirnov@theory.sinp.msu.ru}

\makeatother

\begin{document}
\global\long\def\Sum{\mathop{\Sigma}}
 \abstract{We consider the application of the DRA method to the
case of several master integrals in a given sector. We establish a
connection between the homogeneous part of dimensional recurrence
and maximal unitarity cuts of the corresponding integrals: a maximally
cut master integral appears to be a solution of the homogeneous part
of the dimensional recurrence relation. This observation allows us to make
a necessary step of the DRA method, the construction of the general
solution of the homogeneous equation, which, in this case, is a coupled
system of difference equations.}

\title{The Dimensional Recurrence and Analyticity Method for Multicomponent
Master Integrals:\\
 Using Unitarity Cuts to Construct Homogeneous Solutions}

\maketitle
\allowdisplaybreaks

\section{Introduction}

Recently, a method of evaluating Feynman integrals based on the use
of dimensional recurrence relations \cite{Tarasov1996} and analytic
properties of Feynman integrals as functions of space-time dimension
$d$ (the DRA method) was introduced \cite{Lee2010}. It was successfully
applied in a series of calculations \cite{LeeSmSm2010,Lee2010a,LeeSmSm2010a,LeeTer2010,LeeSmi2010,LeeSmSm2011,Lee2011e}
where master integrals for various families of Feynman integrals were
evaluated exactly in $d$ (in terms of nested sums) and also up
to high order in $\epsilon=2-d/2$ (in terms of the conventional multiple
zeta values (MZV), using {PSLQ}). This fast advance of the
DRA method was partly due to the availability of a number of magnificent
tools and methods: IBP reduction tools, in particular, \texttt{FIRE}
\cite{Smirnov2008}, the sector decomposition analysis of singularities
implemented in \texttt{FIESTA} \cite{SmiSmTe2009}, the application
of Mellin-Barnes technique \cite{Smirnov1999a,Tausk1999a,Czakon2006,Smirnov2006,Smirnov2004,SmirSmi2009a},
the {PSLQ} algorithm \cite{FergBai1991}. The DRA method provides results in the
form of converging (uniformly in $d$) nested sums with factorized
summands. Such a form allows one to evaluate many terms of the $\epsilon$-expansion
with very high precision. This feature of the DRA method was demonstrated
in the evaluation of master integrals for four-loop massless propagators
which were previously evaluated in \cite{BaikChe2010} up to transcendentality
weight seven. Using the results of DRA method it was possible to perform
an evaluation up to weight twelve \cite{Lee2011e}, and it is certainly
possible to go further.

However, up to now, all applications of the DRA method concerned cases
with only one master integral with a given set of denominators (in
a given \emph{sector}). The reason is that the DRA method requires
finding the general solution of the homogeneous part of the dimensional
recurrence relation. For the case of several master integrals in one sector
this problem becomes very nontrivial. The corresponding homogeneous
equation has a matrix form and is equivalent to one difference equation
of order higher than one. One may speculate that this problem is,
in a sense, artificial and the homogeneous system can be decoupled
or, at least, reduced to a triangular form by a proper choice of the
master integrals (i.e., by passing to some linear combinations of
the integrals with rational coefficients). In this case the high-order
difference equation for one master integral should have a hypergeometric-term
solution, which can be checked by the Petkovšek's algorithm {Hyper}
\cite{Petkov1992}. Unfortunately, in real-life examples, the homogeneous
equation appears to have no hypergeometric or d`Alembertian solutions.
So, taken as a separate mathematical problem, finding the solution
of the homogeneous equation for the case of several master integrals
cannot be performed in a systematic way. Therefore, when constructing
the homogeneous solution, one has to rely on some additional methods.
The goal of this paper is to present a method to find the homogeneous
solution using unitarity cuts. The idea is very simple and yet appears
to be very useful.

The paper is organized as follows. In the next Section we introduce
our notation. In Section \ref{sec:Cut-integrals} we show that the
maximal cut of the master integral satisfies the homogeneous part
of the dimensional recurrence relation and this property gives a practical
tool of finding a solution of the homogeneous part of the dimensional
recurrence relations. In Section \ref{sec:A-Three-Loop-Example},
we illustrate our technique on the evaluation of two master integrals
(called $I_{14}$ and $I_{15}$ in \cite{SmiSmSt2010a}) for the three-loop
static quark potential \cite{SmiSmSt2008b,SmiSmSt2010b,SmiSmSt2010a,Anzai:2009tm}.
We reproduce the results presented in \cite{SmiSmSt2010a} and obtained
with the help of the Mellin-Barnes representation \cite{Smirnov1999a,Tausk1999a,Czakon2006,Smirnov2006,Smirnov2004,SmirSmi2009a}
and obtain one more term in $\epsilon$-expansion (weight seven).

\section{General setup}

Let us suppose that we are interested in the evaluation of an $L$-loop
Feynman integral depending on $E$ linearly independent external momenta
$p_{1},\ldots,p_{E}$. There are $N=L(L+1)/2+LE$ scalar products
involving the loop momenta $l_{i}$:

\begin{equation}
s_{ik}=l_{i}\cdot q_{k}\,,\quad i=1,\ldots,L,\quad k=1,\ldots,L+E,
\end{equation}
where $q_{1,\ldots,L}=l_{1,\ldots,L}$, $q_{L+1,\ldots,L+E}=p_{1,\ldots,E}$.
The integral has the form

\begin{equation}
J\left(\nu;\mathbf{n}\right)=\int\frac{d^{d}l_{1}\ldots d^{d}l_{L}}{\pi^{L\mathcal{D}/2}\prod_{\alpha=1}^{N}\left[D_{\alpha}+\epsilon_{\alpha}i\,0\right]^{n_{\alpha}}}\label{eq:LoopIntegral}
\end{equation}
where $\epsilon_{\alpha}=\pm1$ and $\nu=d/2$ is a convenient variable.
The quantities $\epsilon_{\alpha}i\,0=\pm i0$ determine the infinitesimal
shifts of the denominators poles. The scalar functions $D_{\alpha}$
are linear polynomials with respect to $s_{ik}$. The functions $D_{\alpha}$
are assumed to be linearly independent and to form a complete basis
in the sense that any non-zero linear combination of them depends
on the loop momenta, and any $s_{ik}$ can be expressed in terms of
$D_{\alpha}$. The indices $n_{\alpha}$ are assumed to be integer,
and if $n_{\alpha}>0$ we say that the integral has a denominator
$D_{\alpha}$. The integrals having the same set of denominators form
a \emph{sector}.

The dimension shifting relation can be written in two equivalent forms \cite{Lee2010a,Lee2012a}:
\begin{equation}
J\left(\nu-1;\mathbf{n}\right)=\tilde{Q}\left(A_{1},\ldots,A_{N}\right)J\left(\nu;\mathbf{n}\right),\label{eq:dimension-shift up}
\end{equation}
or
\begin{equation}
J\left(\nu+1;\mathbf{n}\right)=\tilde{P}\left(B_{1},\ldots,B_{N}\right)J\left(\nu;\mathbf{n}\right),\label{eq:dimension-shift down}
\end{equation}
where $\tilde{Q}\left(A_{1},\ldots,A_{N}\right)$ and $\tilde{P}\left(B_{1},\ldots,B_{N}\right)$
are some polynomials. The operators $A_{\alpha}$ and $B_{\alpha}$
act as follows:
\begin{align}
A_{\alpha}J\left(\nu;n_{1},\ldots,n_{\alpha},\ldots n_{N}\right) & =n_{\alpha}J\left(\nu;n_{1},\ldots,n_{\alpha}+1,\ldots n_{N}\right),\nonumber \\
B_{\alpha}J\left(\nu;n_{1},\ldots,n_{\alpha},\ldots n_{N}\right) & =J\left(\nu;n_{1},\ldots,n_{\alpha}-1,\ldots n_{N}\right).\label{eq:AB}
\end{align}

In order to obtain the dimensional recurrence relation for some master
integral $J_{1}\left(\nu\right)=J\left(\nu;\mathbf{n}_{1}\right)$, we have to plug it
in Eq. \eqref{eq:dimension-shift up} and reduce the right-hand side
using IBP identities. Observe that the integrals appearing on the
right-hand side of Eq. \eqref{eq:dimension-shift up} belong to the
same sector as $J\left(\nu;\mathbf{n}\right)$ or simpler (lower) sectors. Therefore,
the result of the IBP reduction is also a linear combination of master
integrals belonging to the same, or simpler, sectors. Therefore, if
there are no other master integrals in the same sector as $J_{1}$,
the general form of the dimensional recurrence relation is
\begin{equation}
J_{1}\left(\nu+1\right)=C\left(\nu\right)J_{1}\left(\nu\right)+R\left(\nu\right),\label{eq:DRRsingle}
\end{equation}
where $R\left(\nu\right)$ contains only simpler master integrals,
and $C\left(\nu\right)$ is a rational function. Naturally, the dimensional
recurrence relations for simpler master integrals do not depend on
$J_{1}$. The homogeneous part of this equation can be easily solved
in terms of $\Gamma$-functions. The situation is different if there
is more than one master integral in a given sector. In this case we
will refer to the column of master integrals in a given sector as
a \emph{multicomponent master integral} (MMI). The dimensional recurrence relations for MMI form
a coupled system of equations which can be written in the matrix notation
as
\begin{equation}
\mathbf{J}\left(\nu+1\right)=\mathbb{C}\left(\nu\right)\mathbf{J}\left(\nu\right)+\mathbf{R}\left(\nu\right),\label{eq:DRRmulti}
\end{equation}
where $\mathbf{J}=\left(\begin{array}{c}
J_{1}\\
\vdots\\
J_{k}
\end{array}\right)$ is an MMI and $\mathbb{C}\left(\nu\right)$ is a matrix with rational
elements.

In order to apply the DRA method, we have to find a general solution
of the homogeneous equation $\mathbf{J}_{h}\left(\nu+1\right)=\mathbb{C}\left(\nu\right)\mathbf{J}_{h}\left(\nu\right)$.
This system of difference equations can be reduced to one difference
equation of $k$-th order, for example, for $J_{1,h}$. In particular,
for $k=2$, we have
\begin{equation}
J_{1,h}\left(\nu+2\right)+\tilde{C}_{1}\left(\nu\right)J_{1,h}\left(\nu+1\right)+\tilde{C}_{2}\left(\nu\right)J_{1,h}\left(\nu\right)=0,\label{eq:DRRho}
\end{equation}
where $\tilde{C}_{1,2}\left(\nu\right)$ are rational functions expressed
via matrix elements of $\mathbb{C}\left(\nu\right)$. In general,
difference equations of a high order cannot be solved analytically.
There is, however, a possibility to check whether the equation has
a solution in the form of a hypergeometric term (i.e., such a solution
$f\left(\nu\right)$ that $f\left(\nu+1\right)/f\left(\nu\right)$
is a rational function). This possibility is based on the Petkovšek's
algorithm {Hyper} \cite{Petkov1992}. In fact, the (non-)existence
of a hypergeometric-term solution allows one to claim also the (non-)existence
of a more general solution --- d'Alembertian solution. Unfortunately,
the application of the {Hyper} algorithm to real-life examples
(in particular, to the one considered in Section \ref{sec:A-Three-Loop-Example})
proves that such solutions do not exist. Therefore, solving
homogeneous matrix difference equations is a very nontrivial mathematical problem.

\section{Cut integrals\label{sec:Cut-integrals}}

Let us now improve our notation indicating in the arguments of $J$
also the signs of infinitesimal imaginary shifts in Eq. \eqref{eq:LoopIntegral}:
$J\left(\nu;\mathbf{n}\right)\to J\left(\nu;\mathbf{n};\boldsymbol{\epsilon}\right)$.
Below we omit the first argument $\nu$ where it does not lead to
confusion.

Then, an \textit{\emph{$\alpha$}}\textit{-cut integral} can be defined
as
\begin{equation}
\Delta_{\alpha}J\left(\mathbf{n};\boldsymbol{\epsilon}\right)=J\left(\mathbf{n};\ldots,\epsilon_{\alpha},\ldots\right)-J\left(\mathbf{n};\ldots,-\epsilon_{\alpha},\ldots\right).\label{eq:CutIntegral}
\end{equation}
Similarly, we can define an integral cut over several lines: $\Delta_{\left\{ \alpha_{1},\ldots\alpha_{n}\right\} }J\left(\mathbf{n};\boldsymbol{\epsilon}\right)=\Delta_{\alpha_{1}}\ldots\Delta_{\alpha_{n}}J\left(\mathbf{n};\boldsymbol{\epsilon}\right)$.

For non-positive $n_{\alpha}$ the infinitesimal shifts do not change
the integral, so,
\[
\Delta_{\alpha}J\left(\mathbf{n};\boldsymbol{\epsilon}\right)=0\text{ if }n_{\alpha}\leqslant0\,.
\]
In contrast, for the positive $n_{\alpha}$ the cut integral $\Delta_{\alpha}J\left(\mathbf{n};\boldsymbol{\epsilon}\right)$
is not zero and can be obtained from $J\left(\mathbf{n};\boldsymbol{\epsilon}\right)$
by the replacement
\begin{equation}
\left[D_{\alpha}+\epsilon_{\alpha}i\,0\right]^{-n_{\alpha}}\to\left[D_{\alpha}+\epsilon_{\alpha}i\,0\right]^{-n_{\alpha}}-\left[D_{\alpha}-\epsilon_{\alpha}i\,0\right]^{-n_{\alpha}}=2\pi i\epsilon_{\alpha}\frac{\left(-1\right)^{n_{a}}}{\Gamma\left(n_{\alpha}\right)}\delta^{\left(n_{\alpha}-1\right)}\left(D_{\alpha}\right),\label{eq:cutPrescription}
\end{equation}
where $\delta^{\left(n\right)}\left(x\right)=\frac{d^{n}}{dx^{n}}\delta\left(x\right)$
denotes the $n$-th derivative of Dirac's $\delta$ function. For
$n_{\alpha}=1$ this prescription reduces to the well-known replacement
$\left(D_{\alpha}+i0\right)^{-1}\to-2\pi i\delta\left(D_{\alpha}\right)$.

It is clear that the IBP identities are not sensitive to the specific
choice of $\epsilon_{\alpha}$ in the sense that the coefficients in
these identities do not depend on $\boldsymbol{\epsilon}$. We note,
however, that the symmetry relations are sensitive to this choice
since a symmetry that replaces $D_{\alpha}\to D_{\beta}$ also
replaces $\epsilon_{\alpha}\to\epsilon_{\beta}$. Therefore, temporarily
we consider the master integrals, identical due to the symmetries,
to be different. Then the IBP reduction of an integral $J\left(\mathbf{n}\right)$
is also insensitive to the choice of $\epsilon_{\alpha}$, i.e.
\[
J\left(\mathbf{n},\boldsymbol{\epsilon}\right)=\sum_{i}C^{i}\left(\mathbf{n}\right)J_{i}\left(\boldsymbol{\epsilon}\right),
\]
where only master integrals $J_{i}\left(\boldsymbol{\epsilon}\right)$
depend on the choice of $\boldsymbol{\epsilon}$, but not the coefficients
$C^{i}\left(\mathbf{n}\right)$, which are rational functions of $\mathbf{n}$,
$d$, and external invariants. The master integrals entering the right-hand
side either belong to the same sector as $J\left(\mathbf{n},\boldsymbol{\epsilon}\right)$
or to simpler sectors. Applying the $\Delta_{\alpha}$ operator to
this equation we nullify all integrals without $D_{\alpha}$-denominator.
Thus, cutting all the denominators of $J\left(\mathbf{n},\boldsymbol{\epsilon}\right)$
keeps on the right-hand side only master integrals of the same sector
as $J\left(\mathbf{n},\boldsymbol{\epsilon}\right)$.

The cut integrals are also the basic tool of the powerful
generalized unitarity technique \cite{BeDiDuK1995,BeDiDuK1994} which
provides the possibility to construct scattering amplitudes. In fact,
this strategy of writing an Ansatz as a linear combination of some
basic scalar integrals and constructing the corresponding coefficient
functions is very similar to the strategy of solving IBP relations,
especially within Baikov's method \cite{Baikov1997,Baikov1996a}.

The dimensional recurrence relations are also insensitive to the choice
of $\boldsymbol{\epsilon}$. This is obvious already from the fact
that, at the derivation of this relation, we never needed to specify
explicitly the shifts $\pm i0$ in the denominators. Therefore, restoring
$\boldsymbol{\epsilon}$-dependence in Eq. \eqref{eq:DRRmulti}, we
obtain
\begin{equation}
\mathbf{J}\left(\nu+1;\boldsymbol{\epsilon}\right)=\mathbb{C}\left(\nu\right)\mathbf{J}\left(\nu;\boldsymbol{\epsilon}\right)+\mathbf{R}\left(\nu,\boldsymbol{\epsilon}\right),\label{eq:DRRmulti-1}
\end{equation}
where the inhomogeneous term $\mathbf{R}\left(\nu,\boldsymbol{\epsilon}\right)=\sum_{j}\mathbf{C}^{j}\left(\nu\right)J_{j}\left(\nu;\boldsymbol{\epsilon}\right)$
includes only integrals of the lower sectors. The matrix $\mathbb{C}\left(\nu\right)$,
as well as $\mathbf{C}^{j}\left(\nu\right)$, do no depend on $\boldsymbol{\epsilon}$.
Taking the cut $\Delta_{\left\{ \ldots\right\} }$ over all the denominators
of $\mathbf{J}$, we nullify this term and obtain:
\begin{equation}
\Delta_{\left\{ \ldots\right\} }\mathbf{J}\left(\nu+1;\boldsymbol{\epsilon}\right)=\mathbb{C}\left(\nu\right)\Delta_{\left\{ \ldots\right\} }\mathbf{J}\left(\nu;\boldsymbol{\epsilon}\right).\label{eq:hDRRcut}
\end{equation}
Thus, we arrive at a simple but important observation: \emph{the maximal
cut $\Delta_{\left\{ \ldots\right\} }\mathbf{J}\left(\nu;\boldsymbol{\epsilon}\right)$
of a MMI $\mathbf{J}\left(\nu;\boldsymbol{\epsilon}\right)$ is the
solution of the homogeneous part of the dimensional recurrence relation
for $\mathbf{J}\left(\nu;\boldsymbol{\epsilon}\right)$}.

Two remarks are in order. First, $\delta$-functions in a cut integral
may be too restrictive to give a non-zero result for a specific choice
of the metric signature. This becomes obvious in the Euclidean case,
where the denominators are always positive. But this is also true
for Minkovskian metrics as we will see below. Therefore, to satisfy
all the restrictions imposed by the $\delta$-functions one may have
to choose a more general metric signature $\left(1,1,\ldots,-1,\ldots\right)$.
Second, a cut integral gives only one solution of the difference equation,
while for a $k$-th order equation, there are $k$ independent solutions.
For $k=2$  we can, in principle, find a second solution
in an algorithmic way by the `constant variation' method. However,
we find it possible, and even more convenient, to guess several solutions
of the homogeneous equation by examining the Mellin-Barnes representation
for the cut integral. The guessed solutions can then be checked to
satisfy the homogeneous equation either numerically, or strictly,
by using Zeilberger's method of creative telescoping \cite{Zeilber1990,Zeilber1991}.

\section{A three-loop example\label{sec:A-Three-Loop-Example}}

\begin{figure}
\centering\includegraphics[height=3cm]{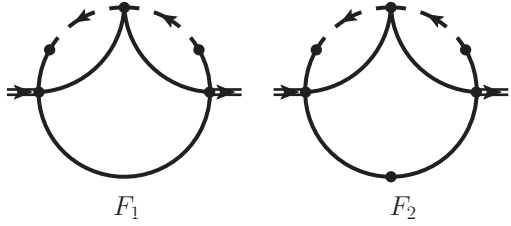}\caption{Master integrals $F_{1}$ and $F_{2}$\label{fig:Master-integrals-F1&F2}}
\end{figure}

\begin{figure}
\centering\includegraphics[height=3cm]{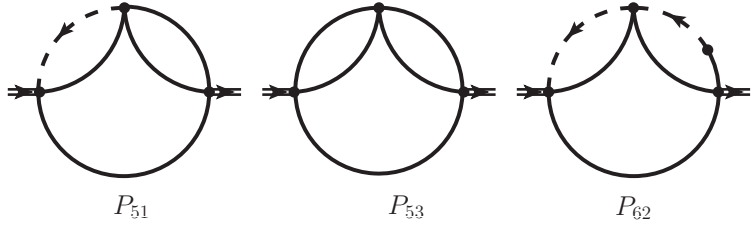}\caption{Simpler master integrals $P_{51},\ P_{53},\ P_{62}$\label{fig:Simler-masters}}
\end{figure}

Let us evaluate the two master integrals shown in Fig.~\ref{fig:Master-integrals-F1&F2}:
\begin{equation}
F_{a}=\int\int\int\frac{\left(i\pi^{d/2}\right)^{-3}\mbox{d}^{d}k\;\mbox{d}^{d}l\;\mbox{d}^{d}r}{(-k^{2})(-r^{2})(-(l+q)^{2})^{a}(-(k-l)^{2})(-(l-r)^{2})(-v\cdot k)(-v\cdot r)}\;,\label{eq:Fa}
\end{equation}
where $a=1$ and $2$, the external momentum $q$ is of the form $(0,\mathbf{q})$,
$v=(1,\mathbf{0})$, and $-i0$ is implied in all the propagators.

The simpler master integrals are depicted in Fig.~\ref{fig:Simler-masters}.
Here we follow the labeling of the master integrals applied in our future 
paper \cite{LeSmSmS}. Moreover, in this labeling, $F_{1}=P_{71}$
and $F_{1}=P_{72}$ but we keep the notation $F_{i}$ which is more
convenient within the present paper. The dimensional recurrence relation
reads:
\begin{equation}
\mathbf{F}\left(\nu+1\right)=\mathbb{C}\left(\nu\right)\mathbf{F}\left(\nu\right)+\mathbf{R}\left(\nu\right),\label{eq:DRR}
\end{equation}
where $\mathbf{F}\left(\nu\right)=\begin{pmatrix}F_{1}\left(\nu\right)\\
F_{2}\left(\nu\right)
\end{pmatrix}$, $\mathbf{R}\left(\nu\right)=\begin{pmatrix}R_{1}\left(\nu\right)\\
R_{2}\left(\nu\right)
\end{pmatrix}$ depends on simpler master integrals, and $\mathbb{C}\left(\nu\right)=\begin{pmatrix}C_{11}\left(\nu\right) & C_{12}\left(\nu\right)\\
C_{21}\left(\nu\right) & C_{22}\left(\nu\right)
\end{pmatrix}$ is a matrix with rational elements. The functions $C_{ij}\left(\nu\right)$
and $\mathbf{R}\left(\nu\right)$ are presented in the Appendix. Observe
that although $C_{ij}\left(\nu\right)$ are quite cumbersome, the
determinant of the matrix $\mathbb{C}\left(\nu\right)$ has a simple
factorized form:
\begin{equation}
\det\mathbb{C}\left(\nu\right)=-\frac{(\nu-2)(4\nu-7)^{2}(4\nu-5)^{2}}{16(\nu-1)^{5}(2\nu-3)^{2}(8\nu-13)(8\nu-11)(8\nu-9)(8\nu-7)}\;.\label{eq:detC}
\end{equation}
This seems to be a general situation.

The homogeneous equation reads
\begin{equation}
\mathbf{F}_{h}\left(\nu+1\right)=\mathbb{C}\left(\nu\right)\mathbf{F}_{h}\left(\nu\right).\label{eq:homDRR}
\end{equation}
The solution of this equation is equivalent to the solution of the
second-order difference equation for $F_{1,h}$
\begin{equation}
F_{1,h}\left(\nu+2\right)+C_{1}\left(\nu\right)F_{1,h}\left(\nu+1\right)+C_{2}\left(\nu\right)F_{1,h}\left(\nu\right)=0\;,\label{eq:homDRR1}
\end{equation}
where $C_{1}$ and $C_{2}$ are known functions. As we already mentioned
earlier, the solution seems to be out of reach of the conventional
mathematical methods based on the use of the {Hyper} algorithm.

In order to apply the DRA method, we need to find two fundamental
solutions of Eq. \eqref{eq:homDRR}, forming a matrix $\mathbb{F}_{h}\left(\nu\right)=\left(\mathbf{F}_{h}^{1}\left(\nu\right),\mathbf{F}_{h}^{2}\left(\nu\right)\right)$.
Then, using the method described in Section 5 of Ref. \cite{Lee2010},
we can find the summing factor $\mathbb{S}\left(\nu\right)$, satisfying
the equation
\begin{equation}
\mathbb{S}\left(\nu\right)=\mathbb{S}\left(\nu+1\right)\mathbb{C}\left(\nu\right).\label{eq:SFeq}
\end{equation}

As we explained in the previous section, the maximally cut MMI $\Delta\mathbf{F}$
satisfies the homogeneous equation \eqref{eq:homDRR}, or, equivalently,
$\Delta F_{1}\left(\nu\right)$ satisfies Eq. \eqref{eq:homDRR1}.
Observe that contracting the lower line of $F_{1}$ in Fig. \ref{fig:Master-integrals-F1&F2}
we obtain a scaleless integral which is zero. Therefore, there is
no need to cut this line as this cut nullifies no simpler master integrals.
In what follows, we also omit the factors $-2\pi i$ from each cut.
Thus, we consider $F_{1}$ and perform the replacements $1/(k^{2}+i0)\to\delta(k^{2})$
and $1/(v\cdot p+i0)\to\delta(v\cdot p)=\delta(p_{0})$ for all the
propagators apart from $1/(-(l+q)^{2})$.

Let us, first, integrate over the loop momenta of the two identical
one-loop subdiagrams consisting of one static and two usual propagators
\begin{equation}
J(l)=\int\frac{\mbox{d}^{d}k}{\pi^{d/2}}\delta(k_{0})\delta(k^{2})\delta(l^{2}-2l\cdot k)\;,\label{delta-int1}
\end{equation}
where $\delta(k_{0})$ comes from $1/(v\cdot k+i0)=1/(k_{0}+i0)$.
Here is a subtle point because in Minkowskian metrics we might conclude
that this integral is zero due to the kinematical restrictions. Indeed,
in Minkowskian space the first two $\delta$-functions result in $k=0$,
which is incompatible with the last $\delta$-function. Let us instead
use the metric signature $\left(1,1,-1,-1,\ldots\right)$, so that
$k^{2}=k_{0}^{2}+k_{1}^{2}-k_{2}^{2}-\ldots-k_{d}^{2}=k_{0}^{2}+k_{1}^{2}-\vec{k}^{2}$.
Then a straightforward integration gives 
\begin{equation}
J(l)=2^{2-d}\frac{\Omega(d-2)}{\pi^{d/2}}\frac{(-l^{2})^{d-4}}{(\mathbf{l}^{2})^{(d-3)/2}}\;,\label{delta-int1-res}
\end{equation}
where $\mathbf{l}^{2}=-l_{1}^{2}+\vec{l}^{2}$, and $\Omega(d)=2\pi^{d/2}/\Gamma(d/2)$
is the volume of the unit hypersphere in Euclidean $d$-dimensional
space. 

To take the final integral
\begin{equation}
\Delta F_{1}\left(\nu\right)=\frac{1}{i^{6}}\int\frac{\mbox{d}^{d}l}{\pi^{d/2}}\frac{J(l)^{2}}{(-(l+q)^{2})}\label{fint}
\end{equation}
we turn to Euclidean space and separate the two terms in the denominator
of $1/(l_{0}^{2}+(\mathbf{l}+\mathbf{q})^{2})$ introducing a onefold
MB representation. The factor $\frac{1}{i^{6}}$ corresponds to six
'time-like' integration variables, two per each loop momenta.

Then the internal integration is taken straightforwardly and we arrive
at the following result:
\begin{align}
\Delta F_{1}\left(\nu\right) & =\frac{2^{4-4\nu}\Gamma\left(6-3\nu\right)}{\Gamma\left(\nu-1\right)^{2}\Gamma\left(8-4\nu,4\nu-\frac{13}{2}\right)}\nonumber \\
 & \times\int\frac{\mbox{d}z}{2\pi i}\frac{\Gamma(-z)\Gamma\left(z+\frac{1}{2}\right)
\Gamma\left(3\nu-\frac{11}{2}-z\right)
\Gamma\left(z-4\nu+8\right)
\Gamma\left(z+\nu-1\right)
 }{\Gamma(z+5-2\nu)}
 \;.\label{eq:MB}
\end{align}
It is easy to convert this representation to a linear combination
of $\,_{3}F_{2}$ hypergeometric functions.

As we mentioned earlier, this gives us only one solution, while a
second-order equation should have two linearly independent solutions.
In order to find both solutions, let us observe that there are two
series of poles from the right of the integration contour and three
series of poles from the left:
\begin{gather*}
z_{1}=n,\quad z_{2}=3\nu-\frac{11}{2}+n,\\
z_{3}=-\frac{1}{2}-n,\quad z_{4}=4\nu-8-n,\quad z_{5}=1-\nu-n,
\end{gather*}
where $n=0,1,\ldots$ It turns out that the contribution of any of
these series constitute a solution of Eq.~\eqref{eq:homDRR1}. This
can be checked either numerically, or using the Zeilberger's method
of creative telescoping \cite{Zeilber1990,Zeilber1991}. We assume,
of course, that the corresponding sums are defined in some region
of $\nu$ where they converge, and then analytically continued to
the whole $\nu$ complex plane. As two independent solutions we choose
the contribution of the series of residues at $z_{1}$ and $z_{4}$.
The solutions have the form
\begin{eqnarray}
F_{1,h}^{1}\left(\nu\right) & = & \frac{\sqrt{\pi}2^{4-4\nu}\Gamma(6-3\nu)\Gamma\left(3\nu-\frac{11}{2}\right)}{\Gamma(5-2\nu)\Gamma(\nu-1)\Gamma\left(4\nu-\frac{13}{2}\right)}\,_{3}F_{2}\left(\left.\begin{array}{c}
8-4\nu,\frac{1}{2},\nu-1\\
5-2\nu,\frac{13}{2}-3\nu
\end{array}\right|1\right)\;,\label{eq:Fh1}\\
F_{1,h}^{2}\left(\nu\right) & = & \frac{32\Gamma(6-3\nu)\Gamma(5\nu-9)\Gamma\left(\frac{5}{2}-\nu\right)}{2^{4\nu}(8\nu-15)\Gamma(\nu-1)^{2}\Gamma(2\nu-3)}\,_{3}F_{2}\left(\left.\begin{array}{c}
8-4\nu,\frac{5}{2}-\nu,4-2\nu\\
10-5\nu,\frac{17}{2}-4\nu
\end{array}\right|1\right)\qquad\label{eq:Fh2}
\end{eqnarray}

Analytical properties of $F_{1,h}^{1}\left(\nu\right)$ and $F_{1,h}^{2}\left(\nu\right)$
can be found from the above representation. Conventional series representation
of the hypergeometric functions $\,_{3}F_{2}$ in Eqs.~\eqref{eq:Fh1},\eqref{eq:Fh2}
converges at $\Re\nu<5/2$. In order to determine the analytical properties
of $F_{1,h}^{1}\left(\nu\right)$ and $F_{1,h}^{2}\left(\nu\right)$
in the region $\Re\nu\geqslant5/2$, one has to use the recurrence
relation \eqref{eq:homDRR1}. It would be more convenient to use the
representation in terms of series converging uniformly in $d$. Luckily,
both $\,_{3}F_{2}$ in Eqs.~\eqref{eq:Fh1} and \eqref{eq:Fh2} appear
to be nearly-poised, and it is possible to transform them to Saalschutzian
$\,_{4}F_{3}$, whose series converge uniformly in $d$. Explicit
expressions of $F_{1,h}^{1}\left(\nu\right)$ and $F_{1,h}^{2}\left(\nu\right)$
in terms of Saalschutzian $\,_{4}F_{3}$ are presented in the Appendix.
Therefore, the fundamental matrix of Eq. \eqref{eq:homDRR} has the
form $\mathbb{F}_{h}\left(\nu\right)=\begin{pmatrix}F_{1,h}^{1}\left(\nu\right) & F_{1,h}^{2}\left(\nu\right)\\
F_{2,h}^{1}\left(\nu\right) & F_{2,h}^{2}\left(\nu\right)
\end{pmatrix},$ where $F_{2,h}^{1}\left(\nu\right)$ and $F_{2,h}^{2}\left(\nu\right)$
are obtained form the first equation of the system \eqref{eq:homDRR}:
\begin{gather}
F_{2,h}^{1}\left(\nu\right)=\frac{F_{1,h}^{1}\left(\nu+1\right)-C_{11}\left(\nu\right)F_{1,h}^{1}\left(\nu\right)}{C_{12}\left(\nu\right)},\quad F_{2,h}^{2}\left(\nu\right)=\frac{F_{1,h}^{2}\left(\nu+1\right)-C_{11}\left(\nu\right)F_{1,h}^{2}\left(\nu\right)}{C_{12}\left(\nu\right)}.\label{eq:F2h}
\end{gather}
Now, following the recipe formulated in Section~5 of Ref.\cite{Lee2010},
we obtain the summing factor
\begin{equation}
\mathbb{S}\left(\nu\right)=\mathbb{W}\left(\nu\right)S\left(\nu\right)\begin{pmatrix}F_{2,h}^{2}\left(\nu\right) & -F_{1,h}^{2}\left(\nu\right)\\
-F_{2,h}^{1}\left(\nu\right) & F_{1,h}^{1}\left(\nu\right)
\end{pmatrix},\label{eq:SF}
\end{equation}
where $S\left(\nu\right)=2^{2\nu}(\nu-2)\Gamma(2\nu-3)^{2}\Gamma\left(4\nu-\frac{13}{2}\right)/\left(\Gamma\left(2\nu-\frac{7}{2}\right)^{2}\Gamma(2-\nu)^{2}\sin(\pi\nu)\right)$
is a solution of the equation $S\left(\nu\right)=S\left(\nu-1\right)\det\mathbb{C}\left(\nu\right)$
and $\mathbb{W}\left(\nu\right)$ is an arbitrary periodic matrix.
Using Eqs.~\eqref{eq:DRR} and \eqref{eq:SF} , we obtain the relation

\begin{equation}
\left(\mathbb{S}\mathbf{F}\right)\left(\nu-1\right)=\left(\mathbb{S}\mathbf{F}\right)\left(\nu\right)+\mathbb{S}\left(\nu-1\right)\mathbf{R}\left(\nu\right),\label{eq:DRRcf}
\end{equation}
which implies
\begin{equation}
\left(\mathbb{S}\mathbf{F}\right)\left(\nu\right)=\mathbf{W}\left(\nu\right)+\Sum_{+\infty}\mathbb{S}\left(\nu-1\right)\mathbf{R}\left(\nu\right),\label{eq:DRRcfsol}
\end{equation}
where $\mathbf{W}\left(\nu\right)$ is an arbitrary periodic column-vector
and the notation $\Sum_{\pm\infty}f\left(\nu\right)$ introduced in
Ref. \cite{Lee2012a} means
\begin{eqnarray}
\Sum_{+\infty}f\left(\nu\right) & = & -\sum_{n=0}^{\infty}f\left(\nu+n\right),\nonumber \\
\Sum_{-\infty}f\left(\nu\right) & = & \sum_{n=1}^{\infty}f\left(\nu-n\right).\label{eq:Spm}
\end{eqnarray}

Now we need to determine $\mathbf{W}\left(\nu\right)$ from the analytical
properties of $\left(\mathbb{S}\mathbf{F}\right)\left(\nu\right)$
which depend on our choice of $\mathbb{W}\left(\nu\right)$. In particular,
if we choose $\mathbb{W}\left(\nu\right)=1$, the function $\left(\mathbb{S}\mathbf{F}\right)$
has singularities at $\nu=2,2\frac{1}{6},2\frac{1}{5},2\frac{1}{3},2\frac{2}{5},2\frac{1}{2},2\frac{3}{5},2\frac{2}{3},2\frac{4}{5},2\frac{5}{6}$
on the stripe $\Re\nu\in\left[2,3\right)$. In order to cancel these
singularities, we can choose $\mathbb{W}\left(\nu\right)$ to be properly
degenerate (and sometimes completely vanishing) matrix at the points
of singularities, but we should also try to not spoil the behavior
of $\left(\mathbb{S}\mathbf{F}\right)$ at $\nu\to\pm i\infty$. Therefore,
it is very useful to eliminate also the explicit and hidden zeros
of $\mathbb{S}$, which, at $\mathbb{W}\left(\nu\right)=1$, are located
at the points $\nu=2\frac{1}{8},2\frac{3}{8},2\frac{5}{8},2\frac{7}{8},\pm i\infty$.
We finally choose
\begin{equation}
\mathbb{W}\left(\nu\right)=\frac{(1+c)(1+2c)}{c^{2}}\left(\begin{array}{cc}
2^{5}(1-c)\left(1-2c-4c^{2}\right) & 2^{5}\frac{1+c}{2c^{2}-1}(1-2c)^{2}\\
-\frac{c}{\sqrt{2}}\left(1-2c-4c^{2}\right) & \frac{c(1-2c)}{\sqrt{2}\left(2c^{2}-1\right)}\left(1-2c-4c^{2}\right)
\end{array}\right),\label{eq:WWef}
\end{equation}
where $c=\cos\left(2\pi\nu\right)$. With this choice of the summing
factor, $\left(\mathbb{S}\mathbf{F}\right)$ is holomorphic in the
stripe $\Re\nu\in\left[2,3\right)$ and grows at $\nu\to\pm i\infty$
slower than $\exp\left(2\pi\left|\nu\right|\right)$. Taking into
account the singularities of $\Sum_{+\infty}\mathbb{S}\left(\nu-1\right)\mathbf{R}\left(\nu\right)$,
we obtain
\begin{equation}
\mathbf{W}\left(\nu\right)=\frac{4\pi^{2}}{\sin^{2}(\pi\nu)}\left(\pi-2\arctan\left(4\sqrt{5}\right)\cos^{2}(\pi\nu)\right)\left(\begin{array}{c}
64\\
-\sqrt{2}
\end{array}\right).\label{eq:Wef}
\end{equation}

Multiplying Eq. \eqref{eq:DRRcfsol} by $\mathbb{S}^{-1}\left(\nu\right)$,
we obtain
\begin{equation}
\mathbf{F}\left(\nu\right)=\mathbb{S}^{-1}\left(\nu\right)\mathbf{W}\left(\nu\right)+\mathbb{S}^{-1}\left(\nu\right)\Sum_{+\infty}\mathbb{S}\left(\nu-1\right)\mathbf{R}\left(\nu\right).\label{eq:DRRres}
\end{equation}
With quantities $\mathbb{S}\left(\nu\right),\ \mathbf{W}\left(\nu\right),\ \mathbf{R}\left(\nu\right)$
determined by Eqs. \eqref{eq:SF},\eqref{eq:WWef},\eqref{eq:Wef},
and \eqref{eq:R}, the above representation \eqref{eq:DRRres} gives
the final result of the DRA method for the MMI $\mathbf{F}\left(\nu\right)=\left(\begin{array}{c}
F_{1}\left(\nu\right)\\
F_{2}\left(\nu\right)
\end{array}\right)$.

Let us make two remarks about the two terms in this representation
of $F_{1,2}\left(\nu\right)$. The second term, in fact, does not
depend on the explicit form of the summing factor $\mathbb{S}\left(\nu\right)$
because
\begin{align*}
\mathbb{S}^{-1}\left(\nu\right)\mathbb{S}\left(\nu+n\right) & =\begin{cases}
\prod_{k=1}^{n}\mathbb{C}\left(\nu+k\right), & n\geqslant0\\
\prod_{k=0}^{-n-1}\mathbb{C}^{-1}\left(\nu-k\right), & n<0
\end{cases}
\end{align*}
is always a finite product of rational matrices. This product can
be evaluated recursively, so that one can organize a numerical evaluation
without nested loops. The first term can explicitly be written as
a combination of fundamental solutions $\mathbf{F}_{h}^{1}$ and $\mathbf{F}_{h}^{2}$:

\[
\mathbb{S}^{-1}\left(\nu\right)\mathbf{W}\left(\nu\right)=\left(\begin{array}{c}
F_{1,h}\left(\nu\right)\\
F_{2,h}\left(\nu\right)
\end{array}\right),
\]
\begin{align}
F_{1,h} & =\frac{2^{5}\pi^{5/2}(2c-1)\left(\pi-(c+1)\arctan\left(4\sqrt{5}\right)\right)}{(2c+1)(1-c)c^{2}}\nonumber \\
 & \times\Biggl[\frac{4c^{3}-2c+1}{2c^{2}-1}F_{1,h}^{1}-\left(4c^2+2c-1\right)F_{1,h}^{2}\Biggr],\nonumber\\
F_{2,h} & =\frac{F_{1,h}\left(\nu+1\right)-C_{11}\left(\nu\right)F_{1,h}\left(\nu\right)}{C_{12}\left(\nu\right)},
\quad c=\cos(2\pi \nu)\,.
\end{align}

Now, taking into account that the evaluation of all the nested sums
appearing in representation \eqref{eq:DRRres} can be organized in
one loop, it is easy to calculate $\mathbf{F}\left(\nu\right)$ with
high precision and apply the {PSLQ} algorithm. Then we obtain:
\begin{multline}
F_{1}\left(2-\epsilon\right)\overset{1000}{=}\frac{28\pi^{4}}{135\epsilon}+\frac{116\pi^{2}\zeta(3)}{9}+\pi^{4}\left(\frac{224}{135}-4\ln(2)\right)+\frac{226\zeta(5)}{3}\\
+\biggl(-192s_{6}+\frac{1808\zeta(5)}{3}-\frac{8\zeta(3)^{2}}{3}+\frac{928\pi^{2}\zeta(3)}{9}+64\pi^{2}\mbox{Li}_{4}\left(\frac{1}{2}\right)+\frac{8}{3}\pi^{2}\ln^{4}(2)\\
-\frac{20}{3}\pi^{4}\ln^{2}(2)-32\pi^{4}\ln(2)-\frac{428\pi^{6}}{2835}+\frac{1792\pi^{4}}{135}\biggr)\epsilon\\
+\biggl(-768\text{Li}_{4}\left(\frac{1}{2}\right)\zeta(3)-128\pi^{2}\text{Li}_{5}\left(\frac{1}{2}\right)+512\pi^{2}\text{Li}_{4}\left(\frac{1}{2}\right)-1536s_{6}+\frac{384}{7}s_{6}\ln(2)\\
-\frac{384s_{7a}}{7}-\frac{3072s_{7b}}{7}+\frac{4960\zeta(7)}{21}+\frac{35519\pi^{2}\zeta(5)}{42}+\frac{14464\zeta(5)}{3}-\frac{64\zeta(3)^{2}}{3}\\
-\frac{31457\pi^{4}\zeta(3)}{945}+\frac{7424\pi^{2}\zeta(3)}{9}-32\zeta(3)\ln^{4}(2)+372\zeta(5)\ln^{2}(2)+32\pi^{2}\zeta(3)\ln^{2}(2)\\
-\frac{480}{7}\zeta(3)^{2}\ln(2)-\frac{3424\pi^{6}}{2835}+\frac{14336\pi^{4}}{135}+\frac{16}{15}\pi^{2}\ln^{5}(2)+\frac{64}{3}\pi^{2}\ln^{4}(2)\\
-\frac{40}{9}\pi^{4}\ln^{3}(2)-\frac{160}{3}\pi^{4}\ln^{2}(2)-\frac{3079}{315}\pi^{6}\ln(2)-256\pi^{4}\ln(2)\biggr)\epsilon^{2}+O(\epsilon^{3})\;,
\end{multline}
\begin{multline}
F_{2}\left(2-\epsilon\right)\overset{1000}{=}-\frac{\pi^{4}}{\epsilon}-93\zeta(5)-14\pi^{2}\zeta(3)-2\pi^{4}\ln(2)\\
+\left(-96s_{6}+120\zeta(3)^{2}+32\pi^{2}\mbox{Li}_{4}\left(\frac{1}{2}\right)+\frac{4}{3}\pi^{2}\ln^{4}(2)-\frac{10}{3}\pi^{4}\ln^{2}(2)-\frac{989\pi^{6}}{420}\right)\epsilon\\
+\biggl(-384\text{Li}_{4}\left(\frac{1}{2}\right)\zeta(3)-64\pi^{2}\text{Li}_{5}\left(\frac{1}{2}\right)+\frac{192}{7}s_{6}\ln(2)-\frac{192s_{7a}}{7}-\frac{1536s_{7b}}{7}\\
-\frac{32666\zeta(7)}{7}-\frac{40585\pi^{2}\zeta(5)}{84}+\frac{35047\pi^{4}\zeta(3)}{630}-16\zeta(3)\ln^{4}(2)+186\zeta(5)\ln^{2}(2)\\
+16\pi^{2}\zeta(3)\ln^{2}(2)-\frac{240}{7}\zeta(3)^{2}\ln(2)+\frac{8}{15}\pi^{2}\ln^{5}(2)-\frac{20}{9}\pi^{4}\ln^{3}(2)-\frac{3079}{630}\pi^{6}\ln(2)\biggr)\epsilon^{2}+O(\epsilon^{3})\;,
\end{multline}
where the notation $\overset{N}{=}$ indicates that the equality holds
numerically with at least $N$ decimal digits,
\begin{eqnarray*}
s_{6} & = & \zeta(-5,-1)+\zeta(6)\;,\\
s_{7a} & = & \zeta(-5,1,1)+\zeta(-6,1)+\zeta(-5,2)+\zeta(-7)\;,\\
s_{7b} & = & \zeta(7)+\zeta(5,2)+\zeta(-6,-1)+\zeta(5,-1,-1)\;,
\end{eqnarray*}
and $\zeta(m_{1},\dots,m_{k})$ are multiple zeta values
\begin{equation}
\zeta(m_{1},\dots,m_{k})=\sum\limits _{i_{1}=1}^{\infty}\sum\limits _{i_{2}=1}^{i_{1}-1}\dots\sum\limits _{i_{k}=1}^{i_{k-1}-1}\prod\limits _{j=1}^{k}\frac{\mbox{sgn}(m_{j})^{i_{j}}}{i_{j}^{|m_{j}|}}\;.\label{MZVdef}
\end{equation}
The terms up to $\epsilon^{1}$ are in agreement with the previous
results \cite{SmiSmSt2010a}.

\section{Conclusion}

We have presented here a method of finding the solution of the homogeneous
part of dimensional recurrence relations for multicomponent master integrals.
The method is based on the fact that the maximally cut master integral
satisfies this homogeneous equation. Strictly speaking, it gives us
only one solution, while for a $k$-component master integral we
need $k$ linearly independent ones. However, it appears that in the
Mellin-Barnes representation of the cut integral each series of poles
separately gives rise to a solution. For each individual case, this
fact can be checked both numerically and strictly, using Zeilberger's
algorithm of creative telescoping.

As an application of this technique, we have presented the
calculation of the two-component master integral $\begin{pmatrix}P_{71}\\
P_{72}
\end{pmatrix}$ given by Eq. \eqref{eq:DRRres} and entering the three-loop static quark
potential. Using this result, we have calculated with a high precision
the $\epsilon$-expansion up to $\epsilon^{2}$-terms and applied
the {PSLQ} algorithm to express it in terms of conventional
constants. Our next natural task is to complete the analytical evaluation
of all the master integrals for the three-loop static quark potential,
i.e. to evaluate the last three analytically unknown expansion coefficients
entering the corresponding result.

\acknowledgments
This work was supported by the Russian Foundation for Basic Research through grant
11-02-01196 and by DFG through SFB/TR~9 ``Computational Particle Physics''.
The work of R.L. was also supported by the Ministry of Education
and Science of the Russian Federation.

\appendix

\section{Coefficients in the dimensional recurrence relation}

Here we present for completeness the quantities $C_{ij}\left(\nu\right)$
and $R_{i}\left(\nu\right)$ entering Eq.~\eqref{eq:DRR}:
\begin{align}
C_{11} & =\!\frac{78656\nu^{6}\!-\!709872\nu^{5}\!+\!2652380\nu^{4}\!-\!5251197\nu^{3}\!+\!5809568\nu^{2}\!-\!3405384\nu\!+\!826308}{4(\nu-1)^{3}(2\nu-3)^{2}(8\nu-13)(8\nu-11)(8\nu-9)(8\nu-7)},\nonumber \\
C_{12} & =
\frac{(\nu -2) \left(13120 \nu ^4-70192 \nu ^3+140108 \nu ^2-123689 \nu +40761\right)}{8 (\nu -1)^3 (2 \nu -3)^2 (8 \nu -13) (8 \nu -11) (8 \nu -9) (8 \nu -7)}\nonumber\\
C_{21} & =\!\frac{3\left(13120\nu^{6}\!-117424\nu^{5}\!+434716\nu^{4}\!-851937\nu^{3}\!+932032\nu^{2}\!-539672\nu\!+129216\right)}{8(1-\nu)^{3}(2\nu-3)^{2}(8\nu-13)(8\nu-11)(8\nu-9)},\nonumber \\
C_{22} &
=-\frac{(\nu -2) \left(6592 \nu ^4-34768 \nu ^3+68324 \nu ^2-59303 \nu +19191\right)}{16 (\nu -1)^3 (2 \nu -3)^2 (8 \nu -13) (8 \nu -11) (8 \nu -9)}.\label{eq:Cs}
\end{align}

\begin{align}
\mathbf{R}\left(\nu\right) & =\begin{pmatrix}R_{1}\left(\nu\right)\\
R_{2}\left(\nu\right)
\end{pmatrix},\nonumber \\
R_{1}\left(\nu\right) & =-\frac{128}{5(8\nu-14)_{8}(3\nu-5)_{3}(2\nu-3)_{2}}\bigl(648214272\nu^{10}-9064230912\nu^{9}\nonumber \\
 & +56911513696\nu^{8}-211292587888\nu^{7}+513701269195\nu^{6}-854608449763\nu^{5}\nonumber \\
 & +985285600699\nu^{4}-777347268382\nu^{3}+401666882358\nu^{2}-122748402735\nu\nonumber \\
 & +16847478900\bigr)P_{51}\left(\nu\right)\nonumber \\
 & -\frac{16(2\nu-3)^{2}}{(8\nu-13)_{7}(3\nu-5)_{3}(\nu-1)}\bigl(1133568\nu^{6}-8922240\nu^{5}+29193664\nu^{4}\nonumber \\
 & -50834923\nu^{3}+49690736\nu^{2}-25855817\nu+5595660\bigr)P_{53}\left(\nu\right)\nonumber \\
 & -\frac{17440\nu^{4}-94208\nu^{3}+189730\nu^{2}-168891\nu+56091}{4(\nu-1)^{3}(4\nu-7)(8\nu-11)(8\nu-9)(8\nu-7)}P_{62}\left(\nu\right),\nonumber \\
R_{2}\left(\nu\right) & =\frac{192}{(8\nu-14)_{7}(3\nu-5)_{3}(2\nu-3)_{2}}\bigl(21590784\nu^{10}-300317184\nu^{9}+1875217824\nu^{8}\nonumber \\
 & -6922120208\nu^{7}+16728915563\nu^{6}-27658192027\nu^{5}+31682083339\nu^{4}\nonumber \\
 & -24828801753\nu^{3}+12740565282\nu^{2}-3865560708\nu+526619520\bigr)P_{51}\left(\nu\right)\nonumber \\
 & +\frac{96(2\nu-3)^{2}}{(8\nu-13)_{6}(3\nu-5)_{3}(\nu-1)}\bigl(47232\nu^{6}-368208\nu^{5}+1192698\nu^{4}-2055044\nu^{3}\nonumber \\
 & +1986779\nu^{2}-1021999\nu+218560\bigr)P_{53}\left(\nu\right)\nonumber \\
 & +\frac{4320\nu^{4}-23000\nu^{3}+45592\nu^{2}-39893\nu+13008}{4(\nu-1)^{3}(4\nu-7)(8\nu-11)(8\nu-9)}P_{62}\left(\nu\right).\label{eq:R}
\end{align}

The simpler master integrals are

\begin{gather}
P_{51}\left(\nu\right)=\frac{\pi^{2}\csc(\pi\nu)\csc(3\pi\nu)\Gamma(\nu-1)^{2}}{\Gamma(5\nu-5)}\;,\label{eq:sMIs}\\
P_{53}\left(\nu\right)=\frac{\pi^{3}\csc^{2}(\pi\nu)\csc(3\pi\nu)\Gamma(\nu-1)^{3}}{\Gamma(4-2\nu)\Gamma(2\nu-2)^{2}\Gamma(4\nu-5)}\;,\nonumber \\
P_{62}\left(\nu\right)=
\frac{\pi ^{5/2} 2^{8-6 \nu } (2 \cos (2 \pi  \nu )-1)   \Gamma \left(\frac{3}{2}-\nu \right) \Gamma \left(2 \nu -\frac{5}{2}\right)}
{(2 \nu - 3) \Gamma \left(4 \nu -\frac{11}{2}\right)\sin ^3(\pi  \nu )\cos (4 \pi  \nu )}
+\frac{\Gamma\left(2\nu-\frac{5}{2}\right)\Gamma\left(\frac{13}{2}-4\nu\right)}{2^{1+6\nu}(2\nu-3)\Gamma\left(\nu-\frac{1}{2}\right)}\nonumber \\
\times\!\Sum_{-\infty}\frac{\left(15536\nu^{5}\!-\!75492\nu^{4}\!+\!144596\nu^{3}\!-\!136177\nu^{2}+62875\nu-11340\right)\Gamma\left(\nu-\frac{1}{2}\right)}{5\cdot4^{-3\nu}(5\nu-4)(5\nu-3)\Gamma\left(2\nu-\frac{1}{2}\right)\Gamma\left(\frac{13}{2}-4\nu\right)}P_{51}\left(\nu\right)\;,\nonumber
\end{gather}
where $\Sum_{-\infty}$ is defined in Eq. \eqref{eq:Spm}. The result for $P_{62}$ presented here is found using DRA method.

\section{Homogeneus solutions via Saalschutzian $\,_{4}F_{3}$ }

Using Eq. (2.4.2.3) from \cite{Slater1966}, we obtain
\begin{align}
F_{1,h}^{1}\left(\nu\right) & =\frac{\pi^{3/2}2^{3-4\nu}\Gamma\left(2\nu-\frac{5}{2}\right)\,_{4}F_{3}\left(\left.\begin{array}{c}
\frac{1}{2},4-2\nu,\frac{9}{2}-2\nu,\nu-1\\
\frac{11}{4}-\nu,3-\nu,\frac{13}{4}-\nu
\end{array}\right|1\right)}{\cos(3\pi\nu)(\nu-2)^{2}\Gamma\left(\frac{11}{2}-2\nu\right)\Gamma\left(\nu-\frac{3}{2}\right)\Gamma(2\nu-4)\Gamma\left(4\nu-\frac{13}{2}\right)}\nonumber \\
 & +\frac{\pi\Gamma(2-\nu)\Gamma\left(2\nu-\frac{5}{2}\right)\,_{4}F_{3}\left(\left.\begin{array}{c}
2-\nu,\frac{5}{2}-\nu,\nu-\frac{3}{2},2\nu-3\\
\frac{3}{4},\frac{5}{4},\nu-1
\end{array}\right|1\right)}{4\cos(3\pi\nu)\Gamma\left(\frac{9}{2}-2\nu\right)\Gamma(\nu-1)^{2}\Gamma\left(4\nu-\frac{13}{2}\right)}\;,\\
F_{1,h}^{2}\left(\nu\right) & =\frac{2^{10\left(\nu-2\right)}\pi\sin(\pi\nu)\Gamma(15-8\nu)\,_{4}F_{3}\left(\left.\begin{array}{c}
4-2\nu,4-2\nu,\frac{9}{2}-2\nu,\frac{5}{2}-\nu\\
\frac{25}{4}-3\nu,\frac{27}{4}-3\nu,3-\nu
\end{array}\right|1\right)}{\sin(5\pi\nu)\Gamma\left(\frac{25}{2}-6\nu\right)\Gamma(2\nu-3)\Gamma(3-\nu)}\nonumber \\
 & +\frac{2^{2\nu-5}\pi\sin(\pi\nu)\Gamma(2-\nu)\,_{4}F_{3}\left(\left.\begin{array}{c}
\frac{1}{2},2-\nu,2-\nu,\frac{5}{2}-\nu\\
\frac{17}{4}-2\nu,\frac{19}{4}-2\nu,\nu-1
\end{array}\right|1\right)}{\sin(5\pi\nu)(8\nu-15)\Gamma\left(\frac{5}{2}-\nu\right)\Gamma(\nu-1)\Gamma(2\nu-3)}\;.
\end{align}

\bibliographystyle{JHEP}

\providecommand{\href}[2]{#2}
\begingroup\raggedright\endgroup

\end{document}